\documentstyle[epsfig]{article}
\title{Crossed-boson exchange contribution \\ and Bethe-Salpeter equation}
\author{
L. Theu{\ss}l\footnote{{\it E-mail address:} 
 lukas@isn.in2p3.fr},
B. Desplanques\footnote{{\it E-mail address:} 
 desplanq@isn.in2p3.fr} \vspace{1cm} \\ 
Institut des Sciences Nucl\'eaires \\
(Unit\'e Mixte de Recherche CNRS-IN2P3, UJF)\\
  F-38026 Grenoble Cedex, France 
 } 
\date{}

\begin{document}

\maketitle

\begin{abstract}
The contribution to the binding energy of a two-body system due to the crossed 
two-boson exchange contribution is calculated, using the Bethe-Salpeter 
equation. This is done for distinguishable, scalar particles interacting via the 
exchange of scalar massive bosons. The sensitivity of the results to the 
off-shell behavior of the operator accounting for this contribution is 
discussed. Large corrections to the Bethe-Salpeter results in the ladder 
approximation are found. For neutral scalar bosons, the mass obtained for the 
two-body system is close to what has been calculated with various forms of the 
instantaneous approximation, including the standard non-relativistic approach. 
The specific character of this result is demonstrated by a calculation involving 
charged bosons, which evidences a quite different pattern. Our results explain 
for some part those obtained by  Nieuwenhuis and Tjon on a different basis. Some 
discrepancy appears with increasing coupling constants, suggesting the existence 
of sizeable contributions involving more than two-boson exchanges.
\end{abstract} 

\section{Introduction}  
The necessity of introducing a contribution due to the crossed-boson exchange 
in the interaction kernel entering the Bethe-Salpeter equation \cite{BETH} has 
been referred to many times in the literature \cite{TODO}. 
This component of the interaction is indeed required if one wants the 
Bethe-Salpeter equation to reproduce results obtained with the Dirac or 
Klein-Gordon equations, which are known to provide a good account of the energy 
spectrum of charged particles in the field of a heavy system (one-body limit). 
At the same time it removes an undesirable contribution of the order $\alpha^3 \, {\rm log} 
\alpha$ (or $\alpha^3$) expected from the simplest ladder approximation 
\cite{FELD}, indirectly supporting the validity of the instantaneous 
approximation in describing the one-boson exchange contribution to the two-body 
interaction. Transparent details about the derivation of the above result are 
scarcely found however and, thus, it may be thought that it is pertinent to QED 
or to the one-body limit. The role of the spin and charge of the exchanged 
boson, as well as its mass, is hardly mentioned.

Recently, Nieuwenhuis and Tjon \cite{NIEU}, employing the Feynman Schwinger 
re\-presentation (FSR), performed a calculation of the energy for a system of 
two equal-mass constituents with zero spin exchanging a zero spin massive boson. 
They found a strong discrepancy with the ladder Bethe-Salpeter approximation, 
pointing to the contribution of crossed-boson exchanges. 
Their binding energies are 
also larger than those obtained from various equations inspired by the 
instantaneous (equal time) approximation. Using a non-relativistic but 
field-theory motivated approach, Amghar and Desplanques \cite{AMGH} found that 
the two-boson exchange contribution to the interaction was cancelled by the 
crossed two-boson exchange one, recovering the instantaneous approximation as an 
effective interaction.  This result was extended to constituents with unequal 
masses but is restricted to spin- and charge-less bosons.

In the present work, we looked at the contribution of the crossed two-boson 
exchange contribution in the framework of the Bethe-Salpeter equation. To the 
best of our knowledge, this is the first time that such a calculation is 
performed. The present study is obviously motivated by the two above works, 
which provide separate benchmarks. In the first case, it may allow one to 
determine which part of the large discrepancy between the FSR results and those 
obtained from the Bethe-Salpeter equation is due to the crossed two-boson 
exchange. In the second case, it is interesting to see whether results of a 
relativistic 
framework confirm the qualitative features evidenced by a non-relativistic 
approach.

The plan of the paper is as follows. In the second section, we derive and 
discuss the expression we will use to describe the contribution of the crossed 
two-boson exchange. A particular attention is given to the off-shell extension 
of the corresponding operator. The third section is devoted to the description 
of the method used to solve the Bethe-Salpeter equation. Results for a system of
two distinguishable, scalar constituents are presented and discussed in the 
fourth section. 

While the present work has not much relationship to the main research acti\-vity 
of W. Gl\"ockle, we don't think it is orthogonal to his physics interests. 
Looking at his work, one can indeed find some publications dealing with 
punctual aspects pertinent to the relativistic description of a few body system. 
We feel that this contribution will nicely fit to his concerns in physics.

\section{Expression of the crossed-boson exchange to the 
 interaction kernel}

\begin{figure}[htb]
\begin{center}
\mbox{\psfig{file=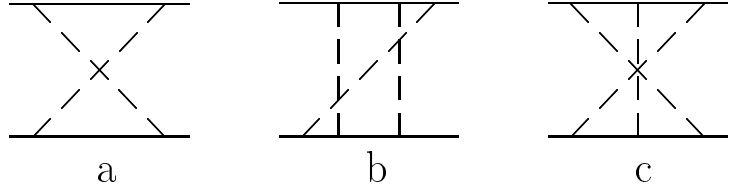,width=30em}}
\caption{\label{fig1}Two- and three crossed-boson exchanges.}
\end{center}
\end{figure}

For definiteness, we first remind the expression of the Bethe-Salpeter equation 
for two scalar particles of equal mass, interacting via the exchange of another
scalar particle. It is given in momentum space by:
\begin{equation}
\label{bseq}
(m^2-(\frac{P}{2}+p)^2)\,(m^2-(\frac{P}{2}-p)^2) \Phi(p,P)
    = i \int \frac{d^4p'}{(2\pi)^4} \, K(p,p',P) \, \Phi(p',P),
\end{equation}
where the quantities $m$, $P$, $p$, $p$' represent the mass of the 
constituents, their total and their
relative momenta, respectively.  $K(p,p',P)$ 
 represents the interaction kernel whose expression is precised 
below. It contains a well determined contribution due to a single boson 
exchange:
\begin{equation}
 K^{(1)} = -\frac{g^{2}}{  \mu^{2} - t} = -\frac{g^{2}}{  \mu^{2} - (p-p')^2 
-i\epsilon},
\label{obe}
\end{equation}
which has been extensively used in the literature. In this equation, the 
coupling, $g^2$, has the dimension of a mass squared. The quantity 
$\frac{g^2}{4m^2}$ is directly comparable to the coupling, $g^2_{MNN}$ 
currently 
used in hadronic physics or to $4\pi\, \alpha$ where $\alpha$ is the usual QED 
coupling. The full fernel $K$ also contains multi-boson exchange contributions 
that  Eq. (\ref{obe}) cannot account for, namely of 
the non-ladder type. Examples of such contributions are shown in Fig. \ref{fig1} 
involving crossed two- and three-boson exchanges.

Dealing with the crossed two-boson exchange should not raise difficulties if 
the 
expression of its contribution was tractable. This one, as derived from the 
Feynman diagram 1a, is quite complicated and, moreover,  not 
amenable to the mathematical methods generally employed in solving Eq. 
(\ref{bseq}). 
There exists however an expression for the on-shell amplitude that has a 
structure similar to Eq. (\ref{obe}). It consists in writing
 a double dispersion relation with respect to 
the $t$ and $u$ variables \cite{MAND}:

\begin{eqnarray}
\label{disprel}
K^{(2)}(s,t,u)=-\frac{g^4}{8\pi^2}\int_{4\mu^2}^{\infty} \int_{4m^2}^{\infty} 
\frac{dt'}{t'-t-i\epsilon} \; \frac{du'}{u'-u-i\epsilon} \;
\frac{1}{  \sqrt{ \kappa(u',t') }  } \; \theta(\kappa(u',t')) , \\
\rule[-1mm]{0mm}{7mm}
{\rm with} \hspace*{1cm} \kappa(u',t')= u'\,t'\,[(t'-4\mu^2)(u'-4m^2)-4\mu^4]. 
\hspace*{18mm} \nonumber
\end{eqnarray}

The quantities $s$, $t$ and $u$ represent the standard Mandelstam variables. For a 
physical process, they verify the equality $s+t+u=4m^2$. The integration 
in Eq. (\ref{disprel}) runs over the domain $t'>4\mu^2$, $u'>4m^2$ 
where  the quantity $\kappa(u',t')$ entering the square root 
function is positive. One of the integrations in Eq. (\ref{disprel}) can be 
performed analytically, allowing one to write:
\begin{eqnarray}
\label{drel2}
K^{(2)}&=&-\frac{g^4}{8\pi^2}\int_{4\mu^2}^{\infty} \frac{dt'}{t'-t-i\epsilon} \; 
\frac{1}{ \sqrt{\kappa(u,t')}}\;  
\frac{1}{2}\; {\rm log}
\left( \frac{ \alpha(u,t')+\frac{1}{2} \sqrt{ \frac{t'-4\mu^2}{t'} }
\sqrt{\kappa(u,t')}   }{ \alpha(u,t')-\frac{1}{2} \sqrt{ \frac{t'-4\mu^2}{t'} }
\sqrt{\kappa(u,t')} } \right), \\ && \nonumber
{\rm with} \hspace*{1cm} \alpha(u,t')=\frac{1}{4} \, [(4m^2- u)(t'-4\mu^2)+ 4 
\mu^4 - u (t'-4\mu^2)] 
 .
\end{eqnarray}

The expression for $\alpha(u,t')$ has been written as the sum of two terms appearing in a 
product form in the second term of the numerator and denominator of the log 
function in Eq. (\ref{drel2}). This suggests an alternative expression for the 
last factor in Eq. (\ref{drel2}) (to be used with care):
\begin{equation}
\frac{1}{2}\;{\rm log}
\left( \frac{ \alpha(u,t')+\frac{1}{2} \sqrt{ \frac{t'-4\mu^2}{t'} }
\sqrt{\kappa(u,t')} 
    }{ \alpha(u,t')-\frac{1}{2} \sqrt{ \frac{t'-4\mu^2}{t'} }
\sqrt{\kappa(u,t')}  } \right)
=  
{\rm log} \left( \frac{ \sqrt{(4m^2-u)(t'-4\mu^2)+4\mu^4} + \sqrt{ 
-u\,(t'-4\mu^2) }}{\sqrt{(4m^2-u)(t'-4\mu^2)+4\mu^4} - 
\sqrt{ 
-u\,(t'-4\mu^2) } } 
\right).
\label{drel3}
\end{equation} 

Expressions for $K^{(2)}$ differing from the previous ones by the front factor 
may be found in the literature (two times smaller in ref. \cite{MAND} and two 
times too large in ref. \cite{ITZY}, which is perhaps due to the 
consideration of
identical particles). For this reason, we felt obliged to enter 
somewhat into details above and precise our inputs.

Methods based on dispersion relations are powerful ones, allowing one to make 
off-shell extrapolations of on-energy shell amplitudes, sometimes far from the 
physical domain. This extrapolation is not unique however (how to extrapolate 
an amplitude which is equal to zero on-energy shell?) and it may thus result 
some uncertainty in a calculation involving a limited number of exchanged 
bosons.

Without considering all possibilities, the dependence on the variable $u$ of the 
crossed two-boson exchange contribution given by Eq. (\ref{drel2}) can provide 
some insight on the above uncertainties. Indeed, the variable $u$, which is an 
independent one in Eq. (\ref{drel2}), can be replaced by $4m^2-s-t$ on-energy 
shell, thus introducing a possible dependence on the variables $s$ and $t$. The 
dependence on $s$ makes the interaction energy dependent. There is nothing wrong 
with this feature but it is not certain that it is physically relevant. It may 
well be an economical way to account for higher order processes. On the other 
hand, the arbitrary character of its contribution should be removed by 
considering the contribution of these higher processes. In any case, the effect 
of the dependence on $s$ gives some order of magnitude for contributions not 
included in this work. The dependence on $t$ through the dependence on $u$ 
raises 
another type of problem. It is illustrated by rewriting the factor appearing in 
Eq. (\ref{disprel}):
\begin{equation}
\frac{1}{(t'-t)(u'-u)}  =  \frac{1}{(t'-t)(u'+s-4m^2+t)} 
  =  \left( \frac{1}{t'-t}+ \frac{1}{u'+s-4m^2+t} \right) \; 
\frac{1}{u'+s-4m^2+t'}.
\label{fact}
\end{equation}
While the first term in the bracket has mathematical properties similar 
to Eq. (\ref{obe}), the second term evidences different properties, poles 
appearing for negative values of $t$. This prevents us from applying the Wick rotation 
and, with it, the numerical methods employed to solve Eq. (\ref{bseq}). 

In the following, we will consider various choices about the off-shell 
extrapolation of Eq. (\ref{drel2}). For non-relativistic systems, we expect the 
uncertainties to be small, in relation with the fact that the quantity $u$ 
appearing in the factor $(u'-u)$ in the denominator of Eq. (\ref{disprel}) is 
small in comparison with $u'$ ($>4m^2$).
     
Our first choice assumes that $u=0$, as it was done in studies relative to the 
contribution of two-pion exchange to the nucleon-nucleon force  
\cite{CHEM}. It is quite conservative and possibly appropriate for a 
non-relativistic approach, which assumes that the potential is energy 
independent. 
The expression of $K^{(2)}$ is then given by:
\begin{equation}
K^{(2)}=- \frac{g^4}{8\pi^2} \int_{4\mu^2}^{\infty} \frac{dt'}{t'-t} 
\; \sqrt{ \frac{t'-4\mu^2}{t'} } \;\; \frac{1}{2( \mu^4 + m^2(t'-4\mu^2))} .
\label{uegal0}
\end{equation}
A second choice consists in replacing the variable $u$ in Eq. (\ref{drel2}) by the
factor $4m^2-s-t'$. Upon inspection, one finds that this is equivalent to
neglecting the second term in the bracket 
of Eq. (\ref{fact}). The corresponding interaction kernel is given by:
\begin{eqnarray}
\label{tegaltp}
K^{(2)} = - \frac{g^4}{8\pi^2} \int_{4\mu^2}^{\infty} \frac{dt'}{t'-t} 
\; \sqrt{ \frac{t'-4\mu^2}{t'} } \;\; \frac{1}{A\,B} \; {\rm log} 
\frac{A+B}{A-B}, \hspace*{2cm} \\
{\rm with}\;\; A=\sqrt{ (t'-2\mu^2)^2+s\, (t'-4\mu^2) }, \;
B= \sqrt{ ( t'+s -4m^2)(t'-4\mu^2) } \nonumber. 
\end{eqnarray}
This expression depends on the variable $s$ and, thus, can provide some order of 
magnitude for the effect due to higher order processes. Its effect can be 
directly  compared to that in the non-relativistic limit corresponding to 
$s=4m^2$, whose expression, also conservative, is given by:
\begin{eqnarray}
\label{segal4}
K^{(2)} = - \frac{g^4}{8\pi^2} \int_{4\mu^2}^{\infty} \frac{dt'}{t'-t} 
\; \sqrt{ \frac{t'-4\mu^2}{t'} } \;\; \frac{1}{A\,B} \; {\rm log} 
\frac{A+B}{A-B}, \hspace*{1.5cm} \\
{\rm with} \;\; A=\sqrt{ (t'-2\mu^2)^2+4 m^2\, (t'-4\mu^2) }, \; 
B= \sqrt{  t'\, (t'-4\mu^2) } \nonumber.
\end{eqnarray}
A last expression of interest corresponds to the non-relativistic limit 
where $m \rightarrow \infty$. Not surprisingly, it is identical to that 
obtained 
by considering time ordered diagrams in the same limit and may be written as:
\begin{equation}
\label{nonrel}
K^{(2)} = - \frac{g^4}{8\pi^2} \int_{4\mu^2}^{\infty} \frac{dt'}{t'-t} 
\; \sqrt{ \frac{t'-4\mu^2}{t'} } \;\; \frac{1}{2 m^2 (t'-4\mu^2)} .
\end{equation}
In this last expression, the possible effect of off-shell extrapolations might 
be studied by replacing the factor $m^2$ by $\frac{s}{4}$.

As a theoretical model, the Bethe-Salpeter equation is most often used with an 
interaction kernel derived from the exchange of a scalar neutral particle. 
Anti\-cipating on the conclusion  that the contribution of the crossed two-boson 
exchange in this case could be strongly  misleading, we also considered the 
case of the exchange of bosons that would carry some ``isospin'' \cite{AMGH}. 
In particular, for the case of two isospin $\frac{1}{2}$ constituents and an
isospin 1 exchange particle, 
a factor $\vec{\tau}_1.\vec{\tau}_2$ would appear in the single 
boson exchange contribution, Eq. (\ref{obe}), while a factor $(3 +2\vec{\tau}_1. 
\vec{\tau}_2)$ should be inserted in the expressions involving the crossed 
two-boson exchange, Eqs. (\ref{uegal0}) - (\ref{nonrel}). 
The factor relative to the 
iterated one-boson exchange would be $(3 -2\vec{\tau}_1.\vec{\tau}_2)$. For a 
state with an ``isospin'' equal to 1, the factor $\vec{\tau}_1.\vec{\tau}_2$ is 
equal to 1 and thus the single boson exchange, Eq. (\ref{obe}), remains 
unchanged. It is easily checked that the factor relative to the ite\-rated single 
boson exchange, $(3 -2\vec{\tau}_1.\vec{\tau}_2)$, is also equal to 1 (as well 
as for the multi-iterated exchanges). On the contrary, for the crossed two-boson 
exchange, the factor $(3 +2\vec{\tau}_1.\vec{\tau}_2)$ makes a strong 
difference. It is equal to 5 instead of 1. Its consequences and the possible 
role of crossed-boson exchange in restoring the validity of the instantaneous 
approximation lost in the ladder Bethe-Salpeter equation will be examined in 
Sect. 4.

\section{Numerical method}

In order to solve Eq. (\ref{bseq}) we employ a variational method 
that, in a similar form, was already used for one of the first
numerical solutions of Bethe-Salpeter equations \cite{SCHW}. 
This method has several advantages, like being little computer time
consuming, and allowing for an easy control of the numerical solutions. 
The drawback lies in the fact, that in the weak binding limit, 
due to our particular choice of trial functions, the convergence of the 
eigenvalues becomes extremely slow, which implies less accurate solutions in 
this limit. This kind of problem is similar to the one encountered in ref. 
\cite{NIU2}, where the rate of convergence was studied in detail.
We have checked carefully, that in the region for which we report numerical
solutions in this work, we reproduce the results of ref. \cite{NIU2} for 
calculations in the ladder approximation.

\subsection{Solution of the equation in ladder approximation}

We consider the problem of two scalar particles of equal mass $m$ 
which interact via the exchange of a third scalar particle of mass $\mu$. After
removal of the center of mass coordinate and performing the Wick rotation to
work in Euclidian space, one is left with an eigenvalue problem in four 
dimensions for the coupling constant $\lambda$. We use the parameter $\epsilon = 
E/2m$, $0<\epsilon<1$, where $E$ is the total energy of the bound state and we 
take $m$ as our mass unit. The Bethe-Salpeter equation in coordinate space takes 
the form

\begin{equation}
L \Psi = \lambda V \Psi,
\label{eq1}
\end{equation}

\noindent
where 

\begin{equation}
L = \left( -\Box +1-\epsilon^2 \right)^2 - 
4 \epsilon^2 \frac{\partial^2}{\partial x_0^2}, \qquad\qquad
 \Box = \sum_{i=1}^{4}\frac{\partial^2}{\partial x_i^2},
\label{eq2}
\end{equation}

\noindent
and the interaction in ladder approximation is 

\begin{equation}
 V(R) = \frac{1}{\pi^2}\int d^4q \frac{e^{-i q \cdot x}}{q^2 + \mu^2} = 
\frac{4 \mu}{R} K_1(\mu R),\qquad\qquad
R = \left( \sum_{i=1}^{4} x_i^2 \right) ^{1/2}.
\label{eq4}
\end{equation}

\noindent
In the last equation, $K_1(x)$ is the modified Bessel function of the 
second kind. For $\mu = 0$, the potential function $V(R)$ becomes simply 
$4/R^2$. The eigenvalue $\lambda$ in Eq. (\ref{eq1}) is related to the
coupling constant $g^2$ used in the previous section by 
$g^2 = (4\pi)^2 m^2 \lambda$.

Equation (\ref{eq1}) is invariant under rotations in the three-dimensional
subspace (but not in the complete four-space except for $\epsilon = 0$). 
One can therefore separate the angular part of the wavefunction, to get a
partial differential equation in two variables. A more convenient way to proceed 
is to switch to spherical coordinates in four dimensions and introduce the 
four-dimensional spherical harmonics

\begin{equation}
|nlm \rangle  = \sqrt{\frac{2^{2l+1}}{\pi} 
\frac{(n+1)(n-l)! \, l!^2}{(n+l+1)!}} \, \,
\sin^l \chi \; \, C_{n-l}^{l+1} \, (\cos \chi) \, \, Y_l^m (\theta, \phi).
\label{eq9}
\end{equation}

\noindent
This choice is particularly suited for our problem, since apart from
obeing simple orthonormality relations, these spherical harmonics
are eigenfunctions of the d'Alembertian operator $\Box$.
The $C_{n}^{l} \, (x)$ in Eq. (\ref{eq9}) are Gegenbauer polynomials.

We expand the functions $\Psi$ of Eq. (\ref{eq1}) in
these four-dimensional spherical harmonic functions, which implies that we
have to determine a set of functions of one variable $R$ only.
We thus arrive at the structure of our trial function

\begin{equation}
\Psi = \sum_n  R^n \, f_{nl}(R)\,\, |nlm \rangle
\label{eq12}
\end{equation}

\noindent
where the factor $R^n$ follows from the asymptotic behaviour of the 
wave function at the origin. Now we expand the functions $f_{nl}$  in terms of 
some basis functions, that are chosen of gaussian type:

\begin{equation}
f_{nl}(R) = \sum_{i} c_i \,\,e^{- \alpha_i R^2}
\label{eq13}
\end{equation}

\noindent
where the $\alpha_i$ are stochastically chosen parameters.  
Note that due to the transformation properties of the spherical harmonic
functions, the sum in Eq. (\ref{eq12}) always runs over either even or odd 
values of $n$, according to the quantum numbers of the state. 
In particular this allows us to calculate states of different relative time 
parity separately.

With the basis functions of Eq. (\ref{eq13}) the matrix elements of $L$ are 
simply computed and for those of $V$ we just need the integrals

\begin{equation}
\int_0^{\infty} R^{2n} e^{-\alpha R^2} K_1(\mu R) \, dR = 
\frac{\Gamma(1+n)\, \Gamma(n)}{2 \alpha^n \mu} \, \, 
e^{\frac{\mu^2}{8 \alpha}} \, \,
W_{-n,\frac{1}{2}} \left( \frac{\mu^2}{4 \alpha} \right),
\label{eq14}
\end{equation}

\noindent
where $W_{\mu,\nu}(z)$ is Whittaker's function.

\subsection{Inclusion of crossed diagrams}

The potential function of Eq. (\ref{eq4}) was obtained from the one-meson
exchange Feynman diagram, which gave a contribution (in Minkowski space and
momentum representation)

\begin{equation}
K^{(1)} = (4\pi)^2 m^2 \frac{\lambda}{t-\mu ^2} ,\;\;\;\;\;  \;\; t=(p - p')^2,
\label{eq15}
\end{equation}

\noindent
where $\mu$ is the mass of the exchanged meson. 
With the results of Sect. 2, the contribution of the crossed
diagram can be expressed with the help of a dispersion relation in the form

\begin{equation}
K^{(2)} = (4\pi)^2 \, 2 m^4 \int_{4\mu^2}^{\infty} 
\frac{\lambda^2}{t-t'} \,\,F(t') \,\,dt',
\label{eq16}
\end{equation}

\noindent
where the spectral function $F(t')$ is given by one of the $t$-independent forms
under the integrals of Eqs. (\ref{uegal0})-(\ref{nonrel}).

Now, since for all the cases considered here,
$F(t')$ is an analytic function eve\-rywhere in the integration domain, we are
allowed to carry out the Wick rotation, and by making the transition to 
configuration space we obtain the generalization of Eq. (\ref{eq1}),

\begin{equation}
L \Psi = \lambda V \Psi + \lambda^2 W \Psi,
\label{eq18}
\end{equation}

\noindent
where the second potential function is given by

\begin{equation}
W(s,R) = \frac{8}{R} \int_{4\mu^2}^{\infty} \sqrt{t'}\, \,
 K_1(R\sqrt{t'}) \, \,F(t') \, \,dt'.
\label{eq19}
\end{equation}

We have indicated in Eq. (\ref{eq19}) by the functional argument that contrary
to V, the new function W may also depend on the total energy of the system, i.e. 
the Mandelstam variable $s$. For the choices of $F(t')$ that we consider here, 
this is only the case for Eq. (\ref{tegaltp}). As we see, Eq. (\ref{eq18}) is no 
longer a simple eigenvalue problem, but it can easily be solved by iteration in 
the form

\begin{equation}
L \Psi = \lambda ( V + \lambda W )\Psi \equiv \lambda U \Psi,
\label{eq20}
\end{equation}

\noindent
where at each step, the determined value of $\lambda$ is reinserted into $U$.

\section{Results}

\subsection{Comparison with non-perturbative results}

We first present results for the ground state mass as a function of the
dimensionless coupling constant $g^2/4 \pi m^2 = 4 \pi \lambda$ for the case 
$\mu / m=0.15$. The reason for this choice is that recently
Nieuwenhuis and Tjon \cite{NIEU} reported the first calculation of
bound state properties beyond the ladder approximation using the Feynman-
Schwinger representation (FSR). Since this formulation takes into account all 
ladder and crossed ladder diagrams, we may expect the results of the present 
approach to lie somewhere between those of the ladder and the Feynman-
Schwinger calculations. If the perturbative
series expansion of the Bethe-Salpeter kernel turns out to converge reasonably
fast, which we would like to be the case, our results should actually be closer
to those of the Feynman-Schwinger approach. This would imply that even higher
order terms of the kernel could safely be neglected in perturbative
calculations.

\begin{figure}[t]
\begin{center}
\mbox{\rule[-0.5cm]{0mm}{3cm}}
\mbox{\psfig{file=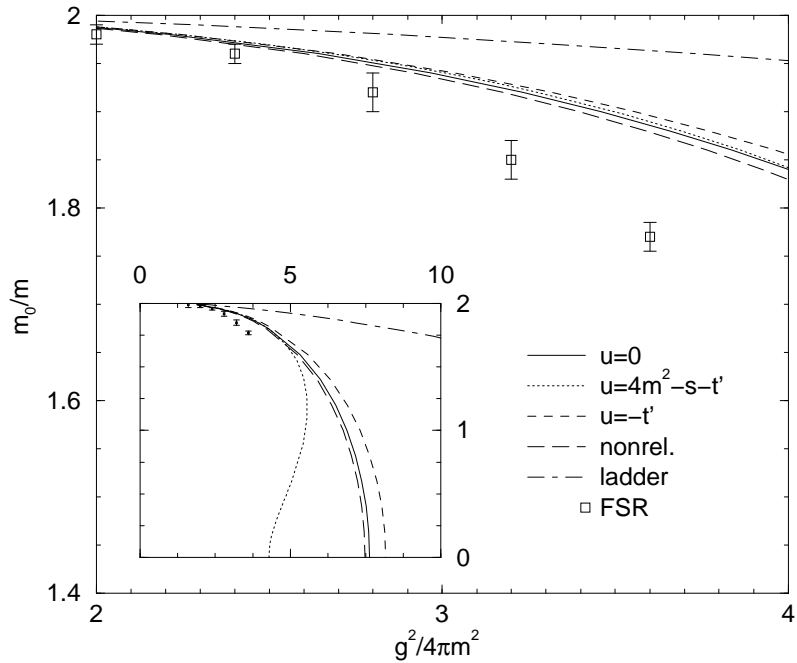,width=30em}}
\end{center}
\vspace{-1cm}
\caption{\label{fig2}
Ground state mass $m_0$ of the scalar theory as a function of the
dimensionless coupling constant $g^2/4\pi m^2$ for $\mu/m=0.15$. The solid,
dotted, dashed and long dashed lines correspond to the solutions with the
spectral functions of Eqs. (\ref{uegal0}) to (\ref{nonrel}) respectively. The
dashed-dotted line gives the ladder result, the boxes represent the
non-perturbative results of Nieuwenhuis and Tjon \cite{NIEU}. The inset shows 
the evolution of the various solutions over the whole range of bound state 
energy values.}
\end{figure}

In Fig. (\ref{fig2}) we show the ladder results and the
Feynman-Schwinger results (taken from ref. \cite{NIEU}) together with the
various off-shell extrapolations given by Eqs. (\ref{uegal0}) - (\ref{nonrel}). 
For small couplings, the differences between these various choices are small 
compared to the difference between the ladder and the Feynman-Schwinger results,
allowing one to make safe statements about the contribution of the crossed 
two-boson exchange diagram. The binding energies so obtained are
found just about halfway between the FSR- and the ladder results. The remaining, 
still considerable, discrepancy to the exact binding energies of the 
non-perturbative approach makes us believe, that even higher order terms than 
the crossed two-boson exchange term in the kernel are essential for doing
reliable calculations within the Bethe-Salpeter framework at large coupling.

The inset of Fig. (\ref{fig2}) shows the evolution of our solutions over the
whole range of binding energies. As can be seen, the solution corresponding to 
the parametrization of Eq. (\ref{tegaltp}) shows a double-valued structure with
a lower branch where the binding energy decreases with increasing coupling. This 
is a consequence of the energy dependence of this choice, since replacing $s$ by 
$4 m^2$, which corresponds to Eq. (\ref{segal4}), renders the curve 
monotonically decreasing. It is however interesting to note that the lower 
branch does not start off at a coupling constant equal to 0 for $s=0$, as it is 
the case, for instance, for the Gross equation, (see below). 
Also the lower branch
solution of the equal mass Klein-Gordon equation goes through 0 for $s=0$.

This qualitative difference to well known facts led us to investigate the
$\mu$ dependence of our results. It was found that, from a value of the
exchanged particles mass of about $\mu \sim 0.5$ on, the curves
become monotonically decreasing even for the choice of Eq. (\ref{tegaltp}). 
On the other hand, for $\mu \rightarrow 0$, the lower
branch tends to $s=0$ for vanishing coupling. 

This could have been immediately expected, since for $\mu=0$ the integral of Eq.
(\ref{tegaltp}) diverges, a reminiscent of the original infrared divergence of
the box diagram. This is somehow a pity, since the case $\mu = 0$, 
in the ladder
approximation, corresponds to the Wick-Cutkosky model, which for $s=0$ admits
analytic solutions in the form $\lambda = N(N+1)$ \cite{WICK}. 
With the crossed box diagram included, all these solutions become degenerate at
$\lambda=0$. We shall not comment any more on this subject, as we believe that
further theoretical work is required in order to correctly account for the
contribution of the crossed box diagram in the case $\mu = 0$.

Nieuwenhuis and Tjon also give a detailed comparison of their results with those
of various quasipotential equations. They considered the
BSLT equation, the equal-time (ET) equation and the Gross equation
\cite{BSLT}.
Generally, these equations
reduce the description from a 4-dimensional to a 3-dimensional
one by making an ansatz for the propagators and the potentials involved.
Specifically, the propagator factor of the Bethe-Salpeter equation in ladder
approximation

\begin{equation}
S^{-1}(p)\Phi (p) = \frac{1}{(2 \pi)^4} \,
\int d^4p' \, V(p-p') \, \Phi (p'),
\label{bse}
\end{equation}

\noindent
(that is Eq. (\ref{bseq}) after the Wick rotation)
is replaced by the following forms in the various approaches 
(compare ref. \cite{NIEU}):

\begin{eqnarray}
S_{\rm QPE}(p)
& \stackrel{\rm BSLT}{=} &  \frac{1}{4\sqrt{\vec{
p}^2+m^2}} \frac{2\pi\:\delta (p_0)}{\vec{p}^2+m^2-\frac{1}{4}s},
\\
& \stackrel{\rm ET}{=} &  \frac{1}{4\sqrt{\vec{
p}^2+m^2}}
\frac{2\pi\:\delta (p_0)}{\vec{p}^2+m^2-\frac{1}{4}s}
 \times\left( 2 - \frac{s}{4(\vec{
p}^2+m^2)}\right)\! ,\label{sqpes} \\
& \stackrel{\rm Gross}{=} &  \frac{1}{4\sqrt{s}\sqrt{\vec{p}^2+m^2}}
\frac{2\pi\:\delta
\left(p_0+\mbox{$\frac{1}{2}$}\sqrt{s}-\sqrt{\vec{
p}^2+m^2}\right)}{\sqrt{\vec{p}^2+m^2}-\frac{1}{2}\sqrt{s}}.
\label{vgros}
\end{eqnarray}

\noindent
In the potential function $V(p-p') = K^{(1)}(p,p',P)$ for the 
BSLT and ET equations, the time component is simply neglected:

\begin{equation}
V(p-p')  =  g^2\frac{1}{(\vec{p}-\vec{p}')^2+\mu^2},
\label{vbare}
\end{equation}

\noindent
whereas for the Gross equation $V$ takes the form

\begin{equation}
V(p-p')  =  g^2\frac{1}{(\vec{p}-\vec{p}')^2-(\omega-\omega')^2+\mu^2},
\label{vbgros}
\end{equation}

\noindent
with $\omega=\sqrt{\vec{p}^2+m^2} - \frac{1}{2}\sqrt{s}$ and 
$\omega'=\sqrt{\vec {p}'^2+m^2} - \frac{1}{2}\sqrt{s}$ .

\medskip

\begin{figure}[t]
\begin{center}
\mbox{\rule[-0.5cm]{0mm}{3cm}}
\mbox{\psfig{file=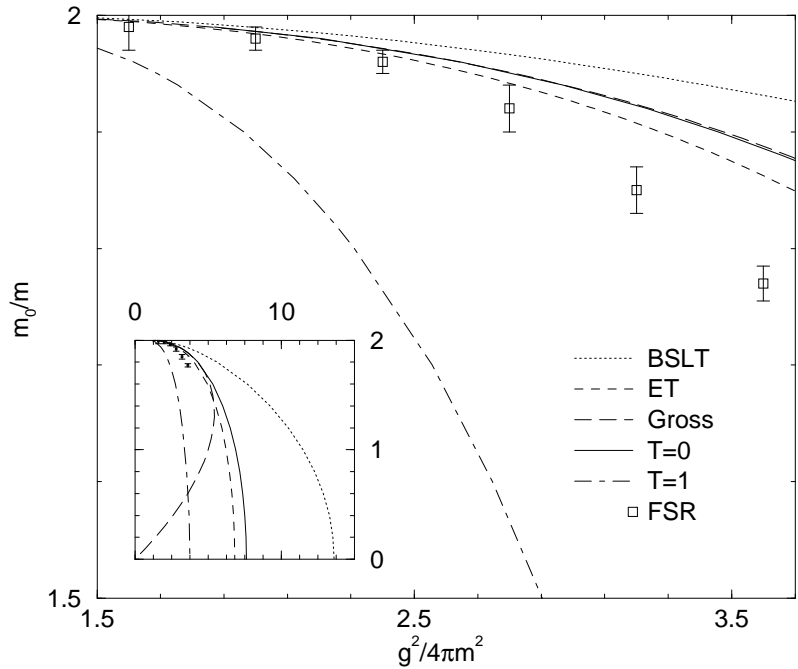,width=30em}}
\end{center}
\vspace{-1cm}
\caption{\label{fig3}
Comparison of ground state masses obtained in various quasipotential equations
with the results of the scalar theory in case of charge-less (solid line) and 
charged (dash-dotted line) particles. The latter one applies for a
state of total isospin 1. Both curves correspond to the choice $u=0$ of Eq.
(\ref{uegal0}).
The boxes represent the non-perturbative Feynman-Schwinger results of 
ref. \cite{NIEU}.}
\end{figure}

In ref. \cite{NIEU} it was found that all the binding energies
obtained in these quasipotential models are distributed between the energies 
of the FSR- and 
the ladder calculations. In Fig. (\ref{fig3}) we compare our results with those
of the equations spe\-ci\-fied above. For small coupling constants, we find that
our binding energies, 
re\-presented by the choice $u=0$ indicated in Fig (\ref{fig3}), are
remarkably close to those of the ET and Gross equation.

 Since it is often conjectured that
results like these may tend to support the validity of the instantaneous
approximation, we would like to point out that this is true only for the
specific model considered here. In fact, one can quickly convince oneself of 
the
special nature of this approximate agreement by considering another specific 
case, where the involved particles would carry some sort of isospin. For the
case already mentioned in Sect. 2, the contribution of
the crossed box for a state of isospin 1 gets multiplied by a factor five, 
whereas the ladder terms remain unchanged. It can be seen in Fig. (\ref{fig3}),
that the corresponding curve even largely overshoots
the Feynman-Schwinger points. The inset of Fig. (\ref{fig3}) shows again the
evolution of the curves over the whole range of binding energies and one can
notice the unphysical lower branch of the Gross equation already mentioned
above and discussed in ref. \cite{NIEU}.

Let us finally notice that results quite 
similar to the ones presented here can also be obtained in a non-relativistic
scheme. Details about this approach will be presented elsewhere \cite{THEU}.

\subsection{Excited states}

\begin{figure}[t]
\begin{center}
\mbox{\rule[-0.5cm]{0mm}{3cm}}
\mbox{\psfig{file=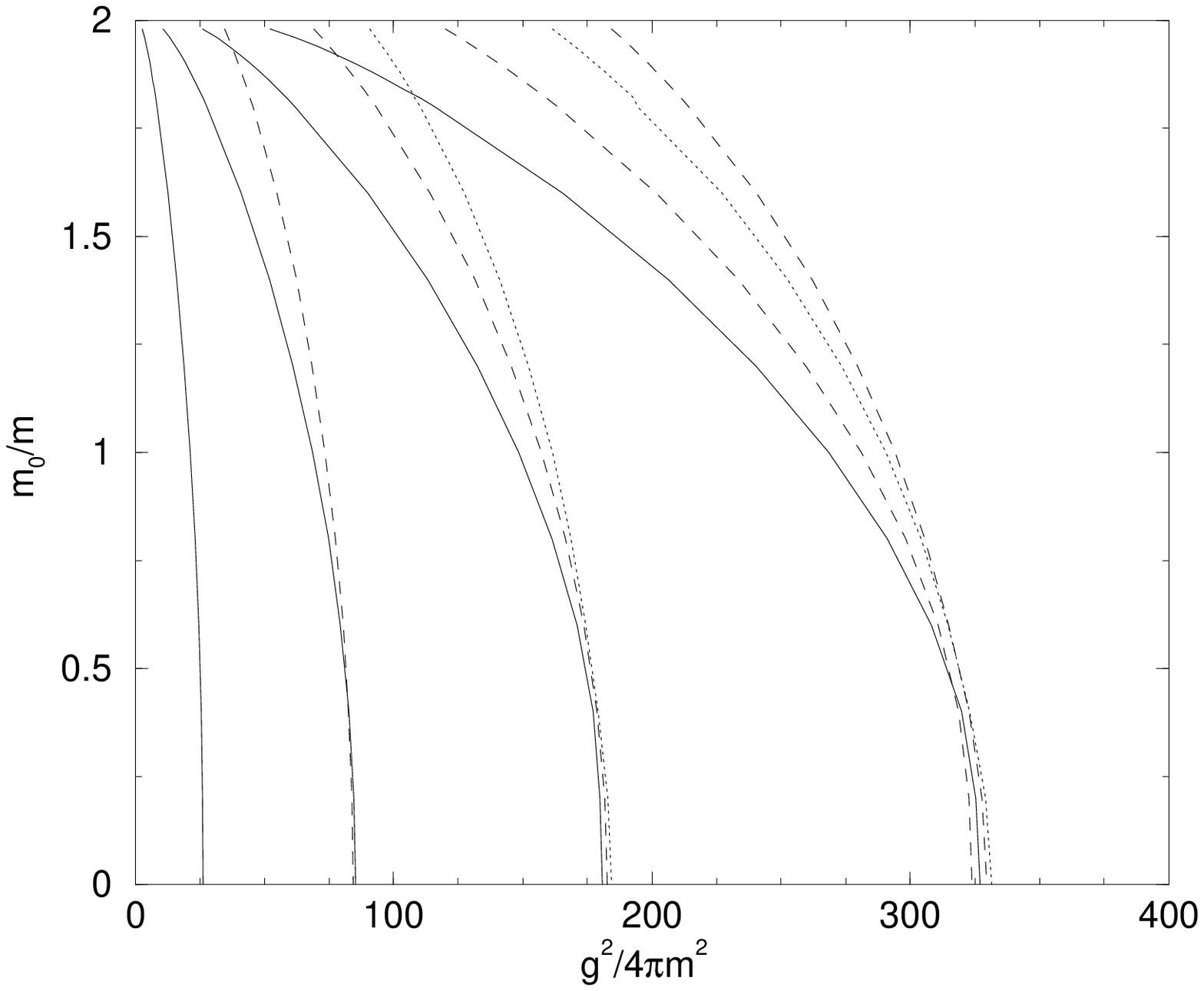,width=16em}
\psfig{file=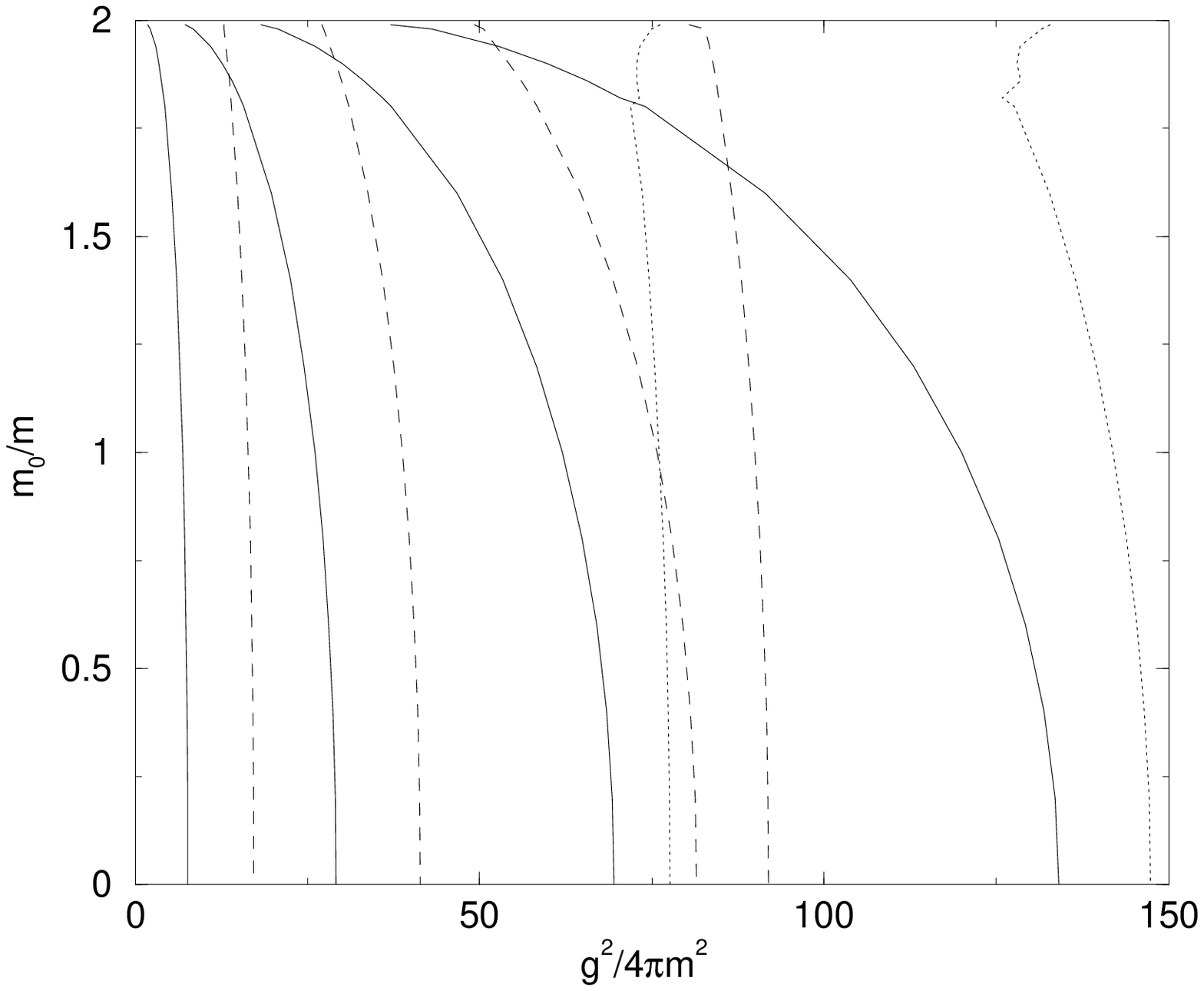,width=16em}}
\end{center}
\vspace{-1cm}
\caption{\label{fig4}
Comparison of the complete spectrum of lowest states of the ladder approximation 
(left) and the one obtained by the inclusion of the crossed box diagram (right) 
for a spatial orbital angular momentum $l=0$. The solid lines give the normal
solutions, dotted and dashed lines give abnormal solutions with even 
and odd time-parity, respectively.}
\end{figure}

In Fig. \ref{fig4} we show  the complete spectrum of lowest states for
a spatial orbital angular momentum $l=0$ and an exchanged particle of mass
$\mu=0.15$. The spectrum of the ladder approximation, on the left hand side, is 
very similar to the one of the Wick-Cutkosky model. There are normal and 
abnormal solutions, the latter ones corresponding to excitations in the relative 
time variable. For $s=0$ an approximate degeneracy appears. This can be traced 
back to the extra symmetry that occurs in the Wick-Cutkosky model, i.e. for 
$\mu = 0$, when $s \rightarrow 0$ (O(5) instead of O(4)). Like in this model, 
the normal states tend to the correct non-relativistic limit when the coupling 
gets small, whereas the abnormal states exist only for larger values of the 
coupling constant. The only qualitative difference to the Wick-Cutkosky model 
is the fact that the different abnormal solutions do not tend to a common value 
of the coupling constant when the binding energy becomes small.

The spectrum gets considerably changed quantitatively by the inclusion of the 
crossed box diagram. On the right hand side of Fig. \ref{fig4} we show the
results for the energy independent choice $u=0$ of Eq. (\ref{uegal0}).
First note the different scale in the coupling constant. Generally, 
the crossed box increases the binding energies for all states as compared with
the ladder results.

Second, it has to be stated that we still find abnormal solutions with in fact 
similar properties as in the ladder approximation. Ever since the initial 
conjecture of  Wick, it has been repeatedly claimed in the literature, that 
these abnormal states should be spurious consequences of the ladder 
approximation. The mere existence of these states beyond the ladder 
approximation is therefore already  an  interesting result per se.

As for the qualitative properties of the solutions, we see that the normal ones
still behave normally, that is, they seem to have the correct weak binding 
limit. Also the abnormal solutions behave like in the ladder approximation in
this limit, since they tend to different but larger values of the coupling
constant. In the strong-binding limit however, the approximate degeneracy 
of states is severely lifted. This could have been expected, since the effect of
the term on the right hand side of Eq. (\ref{uegal0}) is similar to the ladder
term, Eq. (\ref{obe}), with a much higher effective mass $\mu$. An interesting
feature is the fact that the abnormal solutions which are odd functions of
relative time receive a more attractive contribution from the crossed box
diagram than the even time-parity abnormal solutions. This is
clearly seen in Fig. \ref{fig4}, where the odd time-parity solutions get shifted
to the left, the even time-parity solutions to the right of the corresponding
normal solutions. This even leads to the crossing of abnormal states, 
that in the ladder approximation would belong to different O(4) multiplets.

This crossing of states does not cause any numerical problems, since as outlined
in Sect. 3, we calculate even- and odd time-parity solutions separately. There
occurs however a problem for the even time-parity case. Since the normal
solutions all tend to a value that is closer to zero than the limit of
 the abnormal solutions, the former ones necessarily cross some of the latter
ones, if they initially belonged to a higher O(4) multiplet. This leads to a
perturbation of the eigenvalues near the crossing region, which can be clearly
seen from Fig. (\ref{fig4}) in the case of the two positive time-parity
abnormal solutions. The effect is in principle also present in the ladder 
approximation, but it seems somehow less pronounced there. In fact, taking a 
closer look at the crossing region, one finds rather a repulsion of the two 
solutions, with no real crossing taking place. In this region, it is obviously 
not possible to identify the two solutions due to configuration mixing.
It is only by the smooth continuation of the curves after the 
crossing, that the solutions were identified  in order to draw the graph of Fig. 
\ref{fig4}. The curve for the abnormal solution after the crossing is however 
not expected to be meaningfull anymore, since it gets repeatedly perturbed by 
all the normal states crossing it.

\section{Conclusion}
In this work, we have studied the contribution due to crossed two-boson exchange 
to the binding energy of a system made of two distinguishable particles. This 
was done in the framework of the Bethe-Salpeter equation. The present work 
completes that one by Nieuwenhuis and Tjon for the lowest $l=0$ state, allowing 
one to determine what is the role of the simplest crossed-boson exchange 
contribution among all those included in their work. For the range of coupling 
constants where a comparison is possible, this contribution is rather well 
determined and accounts for roughly half of the total effect, the discrepancy 
tending to increase with the coupling. This is consistent with the expectation 
that the role of multi-boson exchanges not included here should increase 
similarly. The result is however somewhat disturbing. It implies that the 
convergence of the Bethe-Salpeter approach in terms of an expansion of the 
interaction kernel in the coupling constant is likely to be slow. We obviously 
assumed that the comparison is meaningful, namely the results of Nieuwenhuis and 
Tjon exclude effects from self-energy or vertex corrections ignored here. 

Amazingly, our results, which include crossed two-boson exchange 
contributions, are close to those obtained with the instantaneous approximation, 
where these contributions are absent. 
This confirms a theoretical result obtained in a non-relativistic 
scheme \cite{AMGH}. Like there, the specific character of this coincidence is 
demonstrated by the consideration of a model with ``charged'' bosons, which 
leads to a different conclusion, showing that the validity of the instantaneous 
approximation is limited to the Born approximation. 

We have looked at excited states, including abnormal ones. Our results are 
partly academic. The effect due to the simplest crossed-boson exchange is so 
large that one has to seriously worry about higher order crossed-boson diagrams. 
Interesting features nevertheless show up. There is no tendancy for abnormal 
states to disappear, as sometimes conjectured in the literature. Their energy is 
differently affected, depending on the parity of the states  under a change in 
the sign of the relative time coordinate.
 
We scarcely considered the case of an exchanged boson with zero mass, 
which would be  most 
interesting in view of possible applications to QED.  This case supposes further 
theoretical work to deal with the divergences that appear. Getting rid of 
the  $\alpha^3 \, {\rm log} \alpha$   corrections to the binding energy obtained 
in the ladder approximation for the case of scalar neutral bosons should be 
easily achieved. Removing the  corrections $\alpha^3$, which are absent in 
results obtained from the Dirac or Klein-Gordon equations, is more delicate. In 
any case, the difficulty has a somewhat general character and is not pertinent 
to the present approach. 

Another development concerns the understanding of the remaining discre\-pancies 
with the FSR results. Estimating the contribution of crossed three-boson  
exchange is not out of reach with some approximations. The net effect is unclear 
however. In the non-relativistic case, at the order $(\frac{1}{m})^0$, it is 
expected to vanish as a result of a cancellation with a contribution due to the 
renormalization of the interaction \cite{AMGH}. It remains the possibility that 
the higher order corrections, which seem to be responsible for a slight extra 
binding in present results, have their role slowly increased when the treatment 
of the problem becomes more complete.


\end{document}